\documentclass[12pt]{article}
=0cm \headsep =0cm

\newcommand{\be}{\begin{equation}}
\newcommand{\ee}{\end{equation}}
\newcommand{\bea}{\begin{eqnarray}}
\newcommand{\eea}{\end{eqnarray}}

\usepackage {latexsym}
\usepackage {graphicx}

\begin{document}

\begin{titlepage}
\title{Asymmetrically Warped Brane Models, Bulk Photons and Lorentz Invariance}

\author{K. Farakos$^a$, N. E. Mavromatos$^b$
 and P. Pasipoularides$^a$ \footnote{paul@central.ntua.gr} \\
       $^a$Department of Physics, National Technical University of
       Athens \\ Zografou Campus, 157 80 Athens, Greece.\\
       $^b$King's College London, University of London\\
       Department of Physics, Strand WC2R 2LS, London, U.K.\\
       }
\date{ }
       \maketitle

\begin{abstract}
We present a brief review of our recent work~\cite{Farakos:2008rv} on asymmetrically warped brane models, where the background metric is characterized by different time and space warp factors.
In particular we examine the case of bulk photons and
we show that the standard Lorentz invariant dispersion relation
is corrected by nonlinear terms which lead to an Energy-dependent speed
of light. Stringent constraints on the parameters of our models can be set by comparing
the results with recent data from high-energy Gamma-Ray Astrophysics, for instance the MAGIC Telescope.
\end{abstract}

\vspace{2cm}
\begin{center}
\emph{Talk given by P. Pasipoularides in NEB XIII 2008, Thessaloniki (Greece) }
\end{center}

\end{titlepage}

\section{Introduction}

Theorists, in an attempt to solve the hierarchy problem, invented new string theory
models with relatively large extra  dimensions. The early realization~\cite{antoniadis}
that the string scale is an arbitrary parameter, and can be as low as a TeV scale, lead naturally to the consideration of models with large extra dimensions ~\cite{ArkaniHamed:1998rs,Antoniadis:1998ig}, introducing the so-called \emph{brane-world models}. In such constructions, the
particle excitations of the Standard Model are assumed to be localized in a 3D
brane (our world), embedded in a multi-dimensional manifold (bulk).
Subsequently, models in which the bulk space time is warped have been proposed~\cite{Randall:1999ee,Randall:1999vf}, and a much richer phenomenology emerged.
In such models, the extra dimensions could be: (i)  finite, if a second parallel brane world lies at a finite bulk distance from our world~\cite{Randall:1999ee}, thus providing a new hierarchy of mass scales,
or: (ii) infinite, if our world is viewed as an isolated brane, embedded in an (infinite) bulk space~\cite{Randall:1999vf}. In fact, it is the presence of such warp factors that provides~\cite{Randall:1999ee} in the case (i) a ``resolution'' to the hierarchy problem.

Beyond the above standard brane world scenarios, there are models
in which all or some of the standard model particles,
can live in the bulk. Note, that in the case of the first Randall-Sundrum (RS) model~\cite{Randall:1999ee} the size of the extra dimension is very small, of the order of the Planck length. In such a case, gauge fields and fermions are not necessarily localized on
the brane, see for example \cite{Davoudiasl:2007wf} and references therein.

There are numerous generalizations of the above generic models, including, for instance, topological defects along the extra dimension(s), or higher-order curvature corrections~\cite{  Mavromatos:2000az, Mavromatos:2002vt, Giovannini:2001ta, Mavromatos:2005yh, BarbosaCendejas:2007hs, Saridakis:2007wx, Zhao}. Also, there are models in which the standard model particles are localized on the brane dynamically, via a mechanism which is known as \emph{layer-phase mechanism}~\cite{Farakos:2005hz,Farakos:2006tt,Farakos:2006sr,Farakos:2007ua,Pasipoularides:2007zz,Dimopoulos:2006qz}. Moreover, an effective propagation of standard model particles in the bulk may characterize the so-called ``fuzzy'' or fluctuating-thick-brane-world scenarios~\cite{campbell}, according to which our brane world
quantum fluctuates in the bulk. In such a case, there are uncertainties in the bulk position of the brane world,
resulting in an ``effectively'' thick brane~\cite{lizzi}.

In this talk we focus on the so called asymmetrically warped brane models.
We shall review briefly work presented in Ref.~\cite{Farakos:2008rv}.
In particular, we consider the following generic ansatz for the metric in five dimensions
\be
ds^2=-\alpha^{2}(z)dt^2+\beta^{2}(z)d\textbf{x}^2+\gamma^2(z) dz^2
\label{asymm}
\ee
where $z$ parameterizes the extra dimension, $\alpha(z)$ is the time warp factor and $\beta(z)$ is
the three-dimensional-space warp factor. Models which are described
by metrics of the form of Eq. (\ref{asymm}), in which the time and three-dimensional-space warp factors are different, are often called asymmetrically warped brane models. In these models, although the induced metric on the brane (localized at $z=0$) is Lorentz invariant
upon considering  the case $\alpha(0)=\beta(0)$,  the metric of Eq. (\ref{asymm}) does not
preserve 4D Lorentz invariance in the bulk since $\alpha(z)\neq\beta(z)$ for $z\neq 0$.

Models with equal warp factors, such as the RS model~\cite{Randall:1999ee,Randall:1999vf},
have so far attracted the main attention, since 4D Lorentz invariance is assumed
as a fundamental symmetry of nature. However, brane models with
asymmetrically warped solutions, like that of Eq. (\ref{asymm}) above, have also been constructed~\cite{Chung:2000ji, Visser:1985qm, Csaki:2000dm, Dubovsky:2001fj, Bowcock:2000cq, Gubser:2008gr, Koroteev:2007yp, Koroteev:2009xd}. Note, that in these models a bulk energy momentum tensor is unavoidable. The question then arises as to how one can constrain or exclude/falsify brane models with \textit{asymmetric}
solutions of the form of Eq. (\ref{asymm}), on account of
present (or immediate-future) experimental bounds on (local) Lorentz symmetry violation in the bulk.

In the standard brane world scenario, where the bulk is completely unaccessible by
the standard model particles, Lorentz violation signals can
be observed only by bulk particles in the gravitational sector,
like gravitons, or at most particles neutral under the standard model group, e.g. right-handed sterile neutrinos.
Bulk fields can "see" the asymmetry between the warp factors in the extra dimension,
whilst standard-model particles, which are rigidly ``pinned'' on the brane,
can only "see" equal warp factors $\alpha(0)=\beta(0)$.
Gravity effects which could reveal 4D Lorentz violation are described
in Ref. \cite{Csaki:2000dm} (see also Refs. \cite{Cline:2001yt, Cline:2002fc, Cline:2003xy}) where superluminous propagation of gravitons is
possible for specific models with \textit{asymmetric} solutions.
Since the detection of gravitons is still not an experimental fact, such Lorentz violations
are still compatible with the current experiments, both terrestrial and astrophysical, probably awaiting the
future detection of gravitational waves in order to be constrained significantly. 

However, in the case where some or all of the
standard model particles, are allowed to propagate in the bulk,
such Lorentz-Invariance-violating effects can be bounded by high precision tests
of Lorentz symmetry, since now, 4D Lorentz violation may
be revealed even in the standard model sector. In this way, stringent restrictions
to \textit{asymmetric} models can be imposed by astrophysical
observations and other high-energy experimental tests. However, even in the case of localized on the brane standard model particles, radiative corrections may induce Lorentz violation operators on the brane, which can set restrictions on asymmetrically warped brane models, see \cite{Frey:2003jq}.

In this paper, we give a brief review of our recent work \cite{Farakos:2008rv}, in which we study the propagation of bulk photons in an
\textit{asymmetrically} warped metric background by
solving the (classical) equations of motion for a 5D massless
U(1) Gauge field in the curved background of Eq. (\ref{asymm}).
We shall demonstrate that the standard Lorentz invariant dispersion relation
for 4D photons possesses nonlinear corrections, which lead to an
energy-dependent speed of light on the brane. Specifically,
we shall obtain a sub-luminal refractive index for photons
$n_{eff}(\omega)=1+c_G\;\omega^2$, where $\omega$ is the energy of
the photon, and the factor $c_G$ is always positive and depends on the free parameters
of the models.
Finally, comparing these results with astrophysical  data, especially from High-energy cosmic Gamma Rays,
we can impose stringent constraints on the parameters of our models. As a concrete example,
we use the recent findings of
the MAGIC Telescope on a delayed arrival
of highly energetic photons from the distant galaxy Mk501~\cite{magic,NM}.

\section{5D U(1) Gauge fields in asymmetrically warped spacetimes}

We consider an action  which includes 5D gravity, a negative cosmological
constant $\Lambda$, plus a bulk U(1) Gauge field~\cite{Csaki:2000dm}:
\begin{equation}
S=\int d^5 x \sqrt{g}\left(\frac{1}{16 \pi G_5}(R^{(5)}-2\Lambda)-
\frac{1}{4}B^{MN}B_{MN} \right)+\int d^4 x \sqrt{g_{(brane)}}{\cal L}_{matter}, \label{matter} \end{equation}
where $G_{5}$ is the five dimensional Newton constant, and $B_{MN}=\partial_{M}H_N-\partial_{N}H_{M}$
is the field strength of the U(1) Gauge field $H_M$, with $M,N = 0,1 \dots 5$.
Note, that this additional bulk Gauge field does not interact with
charged matter on the brane, so it must not be confused with the usual electromagnetic field $A_M$, representing a bulk photon,
which will be introduced later. The four dimensional term
in the action corresponds to matter fields localized on the
brane, which is assumed located at $r=r_0$, and are described by a \emph{perfect fluid}
with energy density $\rho$  and pressure $p$. This brane term is necessary for the
solution of Eq. (\ref{csaki0}) to satisfy the junction conditions on the brane (for details see \cite{Farakos:2008rv} and references there in).

For the metric of the black hole solution we make the ansatz
\begin{equation}
ds^2=- h(r) dt^2+\ell^{-2}r^2d\Sigma^2+h(r)^{-1}dr^2  \label{csaki0}
\end{equation}
where $d\Sigma^2=d\sigma^{2}+\sigma^{2}d\Omega^2$ is the metric of the spatial 3-sections, which
in our case are assumed to have zero curvature, in agreement with the current astrophysical phenomenology, pointing
towards spatial flatness of the observable Universe. Moreover, $\ell$ is the AdS radius which is equal to
$\sqrt{-\frac{6}{\Lambda}}$. By solving the corresponding Einstein equations we obtain:
\begin{equation}
h(r)=\frac{r^2}{\ell^2}-\frac{\mu}{r^2}+\frac{Q^2}{r^4}
\end{equation}
where $\mu$ is the mass (in units of the five dimensional Planck scale) and $Q$ the charge of the 5D \emph{AdS-Reissner-Nordstrom} black hole.

Note that, in the case of nonzero charge $Q$, a non-vanishing component $B_{0r}$ of the bulk field-strength tensor $B_{MN}$:
\begin{equation}
B_{0r}=\frac{\sqrt{6}}{\sqrt{8\pi G_5}}\frac{Q}{r^3}~,
\end{equation}
is necessary in order for the solution to satisfy the pertinent Einstein-Maxwell equations.

In order to achieve our aim, which is to write the 5D AdS-Reissner-Nordstrom
solution as a \emph{linearized perturbation} around the RS metric, we perform
the following change of variables $r \to z(r)$ in Eq. (\ref{csaki0}):
\begin{eqnarray}
r&=& r_0 e^{-k\;z}~, \quad {\rm for}~  z>0 \nonumber \\
r&=& r_0 e^{k\;z}~, \quad  {\rm for}~ z<0~,
\end{eqnarray}
If, in addition,
we make the rescaling $x_{\mu}\rightarrow \frac{r_0}{\ell}x_\mu \quad (\mu=0,\dots 3)$,  we obtain:
\bea
ds^2=-a^2(z) h(z) dt^2+a^2(z)d\textbf{x}^2+h(z)^{-1}dz^2  \label{csaki}
\eea
where $a(z)=e^{-k|z|}$, and $k=\ell^{-1}$ is the inverse $AdS_5$ radius. For the function $h(z)$ we obtain:
\begin{equation}
h(z)=1-\delta h(z), \quad \delta h(z)=\frac{\mu \ell^2}{r_0^4} e^{4 k|z| }-\frac{Q^2 \ell^2}{r_0^6}e^{6k|z|} \label{deviation}
\end{equation}
As we describe in detail in Ref. \cite{Farakos:2008rv} it is not difficult to construct two brane models.
Now, the $Z_2$ symmetry $r\leftrightarrow r_0^2/r$ if it expressed in the frame of the new parameter $z$, reads $z\rightarrow -z$. In addition, the positions of the branes which are located at $r_0$ and $r'_0=r_0 \epsilon$ in the original coordinate system, in the new coordinate system are determined by the equations $z=0$ and $z=\pi r_c$ correspondingly, where $\epsilon=e^{-k\pi r_c}$ and $r_c$ is radius of the compact extra dimension.

Finally, we assume that $|\delta h(z)|\ll 1$ in the interval $0<z<\pi r_c$, or equivalently we adopt that $\delta h(z)$ is
only a small perturbation around the RS-metric. This implies that $r_0$, which is the radius that determines the position
of the brane in the bulk, is (comparatively) a very large quantity. In particular, we have to satisfy both the following two
inequalities $ r_0^2 \epsilon^{2}  \gg \sqrt{\mu} \ell$ and $r_0^3 \epsilon^3 \gg Q \ell$.

Now we will consider the case of a 5D massless $U(1)$ gauge boson $A_{N}$ in the background of an
asymmetrically warped solution of the form of Eq. (\ref{csaki}). This represents a photon propagating in the extra bulk dimensions, which will be of primary interest for our discussion.
The corresponding equation of motion for $A_{N}$ reads:
\be
\frac{1}{\sqrt{g}}\partial_{M}\left(\sqrt{g} g^{MN}g^{RS}F_{NS}\right)=0~, \label{photon}
\ee
with $F_{NS}=\partial_{N}A_S-\partial_{S}A_N$, and $N,S= 0,1,\dots 5$.
In the background metric of Eq. (\ref{csaki}), Eq.~(\ref{photon}) gives:
\bea
-\partial_z(a^{2}(z) h(z) \partial_z A_j)-\nabla^{2}A_j+\frac{1}{h(z)}\partial_{0}^{2}A_j=0, \quad  j=1,2,3~,
\eea
where we have considered the Coulomb gauge condition:
\be
\vec{\nabla }\cdot \vec{A}=0, \quad A_{0}=0,\quad A_{z}=0 ~. \label{coulomp}
\ee
This is suitable for the case of a Lorentz violating background.
On setting in Eq. (\ref{photon}):
\be
A_j(x,z)=e^{i p \cdot x}\chi_j(z), \quad p_{\mu}=(-\omega,\textbf{p})
\ee
and keeping only the linear terms in $\delta h(z)$, we obtain
\be
-\partial_{z}\left\{a^{2}(z) \left[1-\delta h(z)\right]\partial_z\chi\right\} +\left\{\textbf{p}^2-\left[1+\delta h(z)\right]\omega^2\right\}\chi=0 \label{sch1}
\ee
where, for brevity, we have dropped the index $j$ from $\chi$.
Note that the \emph{spectrum} of Eq.~(\ref{sch1}) is \emph{discrete}, due to the orbifold boundary conditions~\cite{Randall:1999ee},
$\chi'(0)=0$ and $\chi'(\pi r_c)=0$ (where the prime denotes a $z$-derivative).

Upon applying the formalism of time-independent perturbation theory to second order (for details see Ref. \cite{Farakos:2008rv}) after same analytical calculations we obtain the following non-trivial dispersion relation for light:
\be
v_{ph}=\frac{\omega}{|\textbf{p}|}=\frac{1}{\sqrt{(1+a_{G})+b_G \omega^2}}
\ee
where $\omega$ is the energy of the photon. For the computation of the parameters $a_G$ and $b_G$ we have obtained the formulas
\be
a_G=\int_{0}^{\pi r_c}dz\left(\chi_{0}^{(0)}(z)\right)^2 \delta h(z) \label{bg0}
\ee
\be
b_G=\sum_{n\neq 0}\frac{1}{\left(m_{n}^{(0)}\right)^2}\left(\int_{0}^{\pi r_c}dz\: \chi_{0}^{(0)}(z)\chi_{n}^{(0)}(z)\delta h(z)\right)^2 \label{bg}
\ee
where $\chi_{n}^{(0)}(z)$ and $m_{n}^{(0)}$ (n=0,1,2..) are the eigenfunctions and eigenvalues of the unperturbed Schrodinger equation for bulk photons, which are known analytically \cite{Davoudiasl:1999tf,Pomarol:1999ad}, see also \cite{Farakos:2008rv}.

If we consider the limit $b_G \omega^2<<1$:
\be
v_{ph}\simeq \frac{1}{\sqrt{(1+a_{G})}}-\frac{b_G}{2(1+a_{G})^{\frac{3}{2}}}\omega^{2} \label{dispr2}
\ee
On the other hand, for the photon's group velocity we have:
\be
v_{gr}\simeq \frac{1}{\sqrt{(1+a_{G})}}-\frac{3 \; b_G}{2(1+a_{G})^{\frac{3}{2}}}\omega^{2} \label{dispr3}
\ee
One can define the constant velocity of light \emph{in standard vacuo} as the low energy limit ($\omega \to 0$) of the phase velocity of Eq. (\ref{dispr2})
\be
c_{light}= \frac{1}{\sqrt{(1+a_{G})}} \label{dispr4}
\ee

From Eqs. (\ref{dispr2}) and (\ref{dispr4}) we then obtain an
\emph{effective subluminal refractive index} for the \emph{non-standard vacuum} provided by our brane-world constructions:
\begin{equation}
n_{eff}(\omega)=\frac{c_{light}}{v_{ph}}=1+\frac{b_G}{2(1+a_G)}\omega^2 ~.\label{effective}
\end{equation}
This is the main result of \cite{Farakos:2008rv}, and of our talk. An issue, we would like to emphasize, is that the phase and group velocities
(Eqs. (\ref{dispr2}) and (\ref{dispr3}) respectively), as well as  the effective refractive index
of Eq. (\ref{effective}), depend quadratically on the photon energy $\omega$.  Moreover, we note that equation (\ref{effective}) is a
perturbative result which is valid only for energies $\omega<<1/\sqrt{b_G}$. For larger energies (for example in the case of ultra high energies cosmic rays $\omega\sim 10^{20}~{\rm eV} $), Eq. (\ref{effective}) is not valid and a full nonperturbative treatment is necessary. In this limit the subluminal nature of the refractive index may be lost and one may also have birefringence effects.
We hope to come to a discussion on such issues in a future work.

\section{Instead of Conclusions: Comparison with Data- the MAGIC observations}

We wish now to compare the time delays of the more energetic photons, implied by the dispersion relation (\ref{effective}),
with experimentally available data. Recently, experimental data from observations of
the MAGIC Telescope~\cite{magic} on photon energies in the TeV range have become available.
It therefore makes sense to compare the time delays predicted in our models, due to (\ref{effective}),
with such data, thereby imposing concrete restrictions on the free parameters of the models.
It should be stressed that the use of the MAGIC data serves only as a concrete example to impose upper bounds on the parameters of our models. From only one set of data one cannot determine with any certainty the physical reasons for the delayed arrival of high energy photons, as compared with their lower-energy counterparts, observed in MAGIC. The source mechanism for the production of high energy photons is still unknown, and certainly one needs many more data and confirmation of certain patterns in the behaviour of high energy photons before reaching conclusions and disentangling source from propagation effects.

For completeness let us first review briefly the
relevant observations~\cite{magic,NM}. MAGIC
is an imaging atmospheric Cherenkov telescope which
can detect very high energy (0.1 TeV-30 TeV) electromagnetic
particles, in particular gamma rays. Photons with very
high energy (VHE) are produced from conversion of
gravitational energy at astrophysical distances from Earth, when, say, a very massive rotating star is
collapsing to form a supermassive black hole. Astrophysical objects
like this are called blazars and are active galactic nuclei (AGN).

The observations of MAGIC during a flare (which lasted twenty minutes)
of the relatively nearby (red-shift $z \sim 0.03$) blazar of Markarian (Mk) 501 on July 9 (2005), indicated a
$4\pm 1$ min time delay between the peaks of the time profile envelops
of photons with energies smaller than 0.25 TeV and photons
with energies larger than 1.2 TeV. Possible interpretations
of such delays of the more energetic photons have already been proposed.
Conventional (astro)physics at the source may be responsible for the delayed emission of
the more energetic photons, as a result, for instance, of some non-trivial versions of the
Synchrotron Self Compton (SSC) mechanism~\cite{magic}. It should be noted at this stage that the
standard SSC mechanism, usually believed responsible for the production of VHE photons in other AGN, such as Crab Nebula, fails~\cite{magic} to explain the results of MAGIC, as a relative inefficient acceleration is needed in order
to explain the ${\cal O}({\rm min})$ time delay.
Modified SSC models have been proposed in this respect~\cite{magic2}, but the situation is not conclusive.
This prompted speculations that new fundamental physics, most likely quantum-gravity effects,
during propagation of photons from the source till observation,  may be responsible for inducing the observed delays, as a result of an effective refractive index for the vacuum, see for example
Ref.~\cite{Ellis:1999sd} and references therein.

Exploring the fact that MAGIC had the ability of observing individual photons, a numerical analysis on the relevant
experimental data has been performed in~\cite{NM}, which aimed at the reconstruction
of the original electromagnetic pulse by maximizing its energy upon the assumption of a sub-luminal vacuum refractive index
with either linear or quadratic quantum-gravity-scale suppression:
\begin{equation}
n_{eff}(\omega)=1+\left(\frac{\omega}{M_{QG(n)}}\right) ^{n}, \; \quad n=1,2 \label{Magic}
\end{equation}
The analysis in \cite{NM} resulted in the following values for the quantum-gravity mass scale at 95 \% C.L.
\begin{equation}
M_{QG(1)}\simeq 0.21 \;~ 10^{18} ~{\rm GeV}, \quad M_{QG(2)}\simeq 0.26 \;~ 10^{11} ~{\rm GeV} \label{Magic0}
\end{equation}
We must emphasize at this point that many Quantum Gravity Models seem to predict
modified dispersion relations for probes induced by vacuum refractive index effects, which appear
to be different for each quantum gravity approach, not only
as far as the order of suppression by the quantum gravity scale is concerned,
but also its super- or sub-luminal nature. Some models, for instance, entail birefringence effects,
which can be severely constrained by astrophysical measurements~\cite{crab}.
There are also alternative models, see for example Refs. \cite{Gogberashvili:2006mr,Carmona:2002iv}.

In \cite{NM1}, a non-perturbative mechanism for the observed time delays has been proposed, based on
stringy uncertainty principles within the framework of a
string/brane theory model of space-time foam. The model entails a brane world crossing regions in bulk space time
punctured by point-like D-brane defects (D-particles). As the brane world moves in the bulk,
populations of D-particles cross the brane, interact with photons, which are attached on the brane world,
and thus  affect their propagation. The important feature of the stringy uncertainty delay mechanism
is that the induced delays are proportional to a single power of the photon energy $\omega$, thus being
linearly suppressed by the string mass scale, playing the quantum gravity scale in this model.
It is important to notice that the mechanism of \cite{NM1} does not entail any modification of the
local (microscopic) dispersion relations of photons. In fact the induced delays are also independent of photon polarization, so
there are no gravitational birefringence effects. It is also important to notice that, in view of
the electric charge conservation, the D-particle-foam defects can interact non trivially only with photons or at most electrically neutral
particles and no charged ones, such as
electrons, to which the foam looks transparent.

As we have showed in the previous section, a
sub-luminal vacuum refractive index may characterize
asymmetrically warped brane-world  models, assuming propagation of photons in the bulk.
Our model, however, predicts a refractive index with quadratic dependence
on energy, c.f. (\ref{effective}).
Comparing Eqs. (\ref{effective}) with (\ref{Magic}), we obtain:
\begin{equation}
\frac{b_{G}}{2(1+a_{G})}\leq M_{QG(2)}^{-2} \label{Magic1}
\end{equation}

The parameters $a_G$ and $b_{G}$ can be computed numerically by means of Eqs. (\ref{bg0}) and (\ref{bg}). Note,
that the deviation $\delta h(z)$, the eigenvalues $m_{n}^{(0)}$ and the eigenfunctions $\chi_{n}^{(0)}$
are known analytically.
We have computed numerically the parameter $b_{G}$ for two exemplary cases: a) for an AdS-Reissner-Nordstrom Solution (\ref{deviation}),
and b) for an AdS-Schwarzschild Solution (obtained by setting $Q=0$ in Eq. (\ref{deviation}))  .
In the former case we assume for simplicity that $\mu$ and $Q^2/r_{0}^2$ are of the same order of magnitude (for details see Ref. \cite{Cline:2003xy} or the discussion in section 2.2 of our work \cite{Farakos:2008rv}),
and hence we can ignore the contribution of the $e^{4 k z}$ term in Eq. (\ref{deviation}).
In particular we obtain:
\begin{eqnarray} &&b_{G}\simeq 2.95\: \langle \delta h  \rangle^{2}\:~ {\rm TeV}^{-2}, \quad {\rm AdS-Reissner-Nordstrom}  \label{Magic3} \\
&&b_{G}\simeq 10 \: \langle \delta h  \rangle^{2}\: ~ {\rm TeV}^{-2}, \quad \:\:\: {\rm AdS-Schwarzschild}  \label{Magic2}~,
\end{eqnarray}
where  $\langle \delta h  \rangle$ is defined as the average value:
\begin{equation}
\langle \delta h  \rangle=\frac{1 }{\pi r_c}\int_{0}^{\pi r_c}\delta h(z) dz~.
\end{equation}
We will use  $\langle \delta h  \rangle$ in order to estimate the degree of violation of Lorentz symmetry in our models.
Taking into account Eqs. (\ref{Magic0}), (\ref{Magic1}) ,(\ref{Magic3}) and (\ref{Magic2}) we find the constraints:
\begin{eqnarray}
&&\mid\langle \delta h  \rangle\mid \leq 1.4 \;10^{-8} \;\quad {\rm AdS-Reissner-Nordstrom}  \label{Magic5} \\
&&\mid\langle \delta h  \rangle\mid \leq 0.75 \;10^{-8} \quad {\rm AdS-Schwarzschild}  \label{Magic4}~.
\end{eqnarray}
The small values we obtain are consistent with the weak nature of $\langle \delta h  \rangle$, as required
by treating it as a perturbation. Also we observe that the values for the average deviation of Eqs. (\ref{Magic5}) and (\ref{Magic4})
are of the same order of magnitude for both AdS-Schwarzschild and AdS-Reissner-Nordstrom solutions.
In the above analysis we ignored the effects due to the Universe expansion, since the latter do not affect the order of magnitude
of the above bounds due to the small red shift ($z \sim 0.03$) of Mk501 we restrict our discussion in this section.
The inclusion of such effects, which are essential for larger redshifts,
is straightforward~\cite{NM} and does not present
any conceptual difficulty.

Note, that although the parameter $a_G$ is not important for our analysis, it is crucial when we have to make comparisons with the velocities of other particles.
If we take into account Eq. (\ref{bg0})  we see that $a_G=\langle \delta h\rangle $. Hence, the parameter $a_G$ is constrained via the equation
\begin{equation}
\mid a_G \mid\leq10^{-8}
\end{equation}
This summarizes the constraints to the asymmetrically-warped brane models with bulk photons using the data of the MAGIC experiment.
For comparison with other data (for example, recent data from H.E.S.S. experiment \cite{hess} and Ultra High Energy Cosmic Rays \cite{sigl}), we refer the reader to our article~\cite{Farakos:2008rv} for further details. Moreover, in that work we have also made some
comparison with the velocity of gravitons (for this latter point, see also \cite{Moore:2001bv}).

The induced modification of the refractive index for photons is a result of the breakdown of the higher(five, in our case) -dimensional Lorentz Invariance and photon propagation in the bulk.
On the brane world, Lorentz-invariant metrics are assumed, but the asymmetric warp factors in the bulk result in feeding the effects of bulk Lorentz-symmetry breaking back onto our brane world, via the induced refractive indices of four dimensional photons.

\section*{Acknowledgements}

P.P. wishes to thank the organisers of NEB XIII 2008 (Thessaloniki (Greece)) for the opportunity to present the results of this work, and for providing a stimulating atmosphere during the meeting.
The work of N.E.M. is partially supported by the European Union through the FP6 Marie-Curie Research and Training Network \emph{Universenet}
(MRTN-2006-035863).

\end{document}